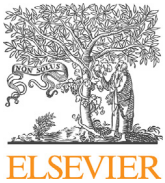
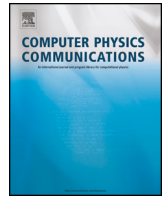

# QOptCraft: A Python package for the design and study of linear optical quantum systems ☆,☆☆

Daniel Gómez Aguado [a], Vicent Gimeno [b], Julio José Moyano-Fernández [b], Juan Carlos Garcia-Escartin [c],*

[a] *Universidad de Valladolid, Valladolid, Spain*
[b] *Universitat Jaume I, Campus de Riu Sec, Departament de Matemàtiques & Institut Universitari de Matemàtiques i Aplicacions de Castelló–IMAC, 12071, Castellón de la Plana, Spain*
[c] *Departamento de Teoría de la Señal y Comunicaciones e Ingeniería Telemática, ETSI de Telecomunicación, Universidad de Valladolid, Campus Miguel Delibes, Paseo Belén 15, 47011 Valladolid, Spain*

## A B S T R A C T

The manipulation of the quantum states of light in linear optical systems has multiple applications in quantum optics and quantum computation. The package QOptCraft gives a collection of methods to solve some of the most usual problems when designing quantum experiments with linear interferometers. The methods include functions that compute the quantum evolution matrix for $n$ photons from the classical description of the system and inverse methods that, for any desired quantum evolution, will either give the complete description of the experimental system that realizes that unitary evolution or, when this is impossible, the complete description of the linear system which approximates the desired unitary with a locally minimal error. The functions in the package include implementations of different known decompositions that translate the classical scattering matrix of a linear system into a list of beam splitters and phase shifters and methods to compute the effective Hamiltonian that describes the quantum evolution of states with $n$ photons. The package is completed with routines for useful tasks like generating random linear optical systems, computing matrix logarithms, and quantum state entanglement measurement via metrics such as the Schmidt rank. The routines are chosen to avoid usual numerical problems when dealing with the unitary matrices that appear in the description of linear systems.

**Program summary**
*Program Title:* QOptCraft
*CPC Library link to program files:* https://doi.org/10.17632/r24hszggf4.1
*Developer's repository link:* https://github.tel.uva.es/juagar/qoptcraft
*Licensing provisions:* Apache-2.0
*Programming language:* Python 3 v3.9.5:0a7dcbd, May 3 2021 17:27:52
*Supplementary material::* User's manual at https://github.tel.uva.es/juagar/qoptcraft/-/blob/main/QOptCraft_V1.1_user_guide.pdf
*Nature of problem:* The evolution of the quantum states of light in linear optical devices can be computed from the scattering matrix of the system using a few alternative points of view. Apart from being able to compute the evolution through a known optical system, it is interesting to consider the less studied inverse problem of design: finding the optical system which gives or approximates a desired evolution. Linear optical systems are limited and can only provide a small subset of all the physically possible quantum transformations on multiple photons. Choosing the best approximation for the evolutions that cannot be achieved with linear optics is not trivial.
This software deals with the analysis of the quantum evolution of multiple photons in linear optical devices and the design of optical setups that achieve or approximate a desired quantum evolution.
*Solution method:* We have automated multiple computation processes regarding quantum experiments via linear optic devices. The methods rely on the properties of the groups and algebras that describe the problem of light evolution in linear system.

---

The library `QOptCraft` for Python 3 includes known numerical methods for decomposing an optical system into beam splitters and phase shifters and methods to give the quantum evolution of system classically described by a scattering matrix using either the Heisenberg picture evolution of the states, a description based on the permanents of certain matrices or the evolution from the effective Hamiltonian of the system. It also provides methods for the design of achievable evolutions, using the adjoint representation, and for approximating quantum evolutions outside the reach of linear optics with an iterative method using Toponogov's comparison theorem from differential geometry. The package is completed with useful functions that deal with systems including losses and gain, described with quasiunitary matrices, the generation of random matrices and stable implementations of the matrix logarithm.

*Additional comments including restrictions and unusual features:* The package is designed to work with intermediate scale optical systems. Due to the combinatorial growth in the state space with the number of photons and modes involved, there is an upper limit on the efficiency of any classical calculation. `QOptCraft` serves as a design tool to explore the building blocks of photonic quantum computers, optical systems that generate useful quantum states of light for their use in metrology or other applications or the design of quantum optics experiments to probe the foundations of quantum mechanics.

## 1. Introduction: linear optical quantum systems

The evolution of the quantum states of light inside linear optical systems shows a rich structure and has applications ranging from fundamental quantum optics experiments to the preparation of advanced quantum states for quantum information processing.

The package `QOptCraft` offers different functions that help to automate the design process for quantum optical experiments. The object under study are linear optical multiports, or linear interferometers. We consider linear interferometers acting on $m$ separate modes and their action on $n$ input photons.

The main purpose of the package is exploring which evolutions can be realized with linear optics and which cannot and giving a physical implementation for those evolutions which are possible. The code is written in Python due to its simplicity, its compatibility with multiple platforms and the existence of many scientific libraries which can be combined with our package. The package offers a library of functions which can be used as a starting point for independent code development as closed, black-box functions for the users that just need to know the experimental setup for their desired evolution.

While the code is fully functional and some tweaks for efficiency have been applied, it is not optimized for computational speed. The combinatorial growth in the state space in the problem makes any implementation in systems with more than a few tens of photons and modes impractical for any computer. The package has been created for exploring medium-scale optical experiments which can be performed before noise or other imperfections become too problematic. The possible applications include the design of optical systems for the generation of entangled states of light, the prediction of the behaviour of medium-sized linear interferometers or to build simple gates for specific quantum information protocols.

This paper serves as a guide to the package and the theoretical foundations on which it was built. It is intended to be a brief tour on the mathematical and physical results that are needed to describe linear interferometers with quantum inputs. The detailed user guide with a functional description of the software is given with the package [1].

The paper starts with a brief description of different existing software and computer design methodologies for quantum linear optical systems and a comparison to our approach (Section 2). The theory behind the calculations in the package is introduced with a review of the matrices that describe linear optical systems (Section 3) and an explanation of all the relevant algebraic objects that appear in the study (Section 4). Section 5 gives a tour on the different methods offered in the package. Section 6 gives complete design examples using the package. The paper ends with a summary (Section 7).

## 2. Comparison to similar existing libraries and other computer-assisted design methods for quantum linear optics

There are many complementary ways of studying the behaviour of the quantum states of light. In `QOptCraft`, the point of view is centered on the unitary evolution matrices that describe the evolution of photon number states with a constant number of photons and modes.

For certain tasks, our functions sometimes overlap with existing software packages and, in many cases, provide an alternative way of looking at things. In this Section, we discuss the most relevant software packages and computer design methods for quantum optical evolutions.

`QuTiP` is a Python framework for the simulation of open quantum systems [2,3], including quantum optical processes. In particular, it offers tools to simulate the evolution for Hamiltonians involving light-matter interaction. It includes examples that describe simple linear optical systems, but the focus of the framework tends to be on time-dependent evolution and the coupling with matter.

Xanadu's `Pennylane` has two wrappers that deal with linear optics. `Strawberry fields` [4] gives a collection of functions that permit to work with linear optical circuits, similar to the usual quantum circuit model of quantum computing.

The whole collection has Continuous Variables, CV, quantum computing [5–7] in mind and, as such, focus mostly on Gaussian operations, homodyne measurement and coherent and squeezed states. There is also a wrapper, `the Walrus` [8], devoted to the study of Gaussian boson sampling [9–11]. This library can be combined with machine learning to design different quantum evolutions [12,13].

This is more similar to our approach of treating linear optical systems as blocks with time independent calculations. However `QOptCraft` works with finite state spaces generated from a discrete constant number of modes and photons instead of the continu-ous variable approach of Strawberry fields with coherent and squeezed states. These states live in an infinite-dimensional space and are described either by operators or truncated state spaces.

In addition to the simulation of known optical systems, there are multiple computer-assisted methods to search for optical setups that produce certain transformations. In the last few years there has been an explosion in the use of machine learning to understand and design physical systems. Some of the possible applications are exploring the foundations of quantum mechanics and the design of nanophotonic structures (see [14] and [15], respectively, for the corresponding surveys). We will give a brief description of the results which are closer to our software. The interested reader can find a complete review of the use of computers to find new quantum optical experiments in [16].

In most of the cases, the purpose of computer-assisted methods is to produce certain resource states, such as highly entangled photon states. The usual approach is starting with simple, easy to generate states and, then, the computer tries different optical systems with elements that are optimized so that the output state is close to some desired state or has a good value for a particular objective function.

The AdaQuantum software [17] builds on the work of Knott [18] and searches for specific output quantum states exploring the configuration space of the optical systems either with genetic algorithms alone [19] or combined with neural networks [20]. These methods are focused on Gaussian states.

The functions in QOptCraft work with photon number states. This is closer to the work found in the papers from Krenn's group, which expand the original proposal of the MELVIN algorithm in [21] to look for new experiments [22,16] with discrete photon states. The MELVIN algorithm searches random linear optical transformations produced from simple elements and performs advanced numerical optimization to tune the devices in order to obtain output states with high scores in an objective function, like, for instance, a measurement of entanglement. This algorithm and has led to multiple successful experiments [23–25]. Interestingly, a second algorithm, THESEUS, can be used to gain insight into new methods. The representation of the problem is designed for easy human interpretation and it aids to the conceptual understanding of the results [26,27]. The outputs from this algorithm have been later refined by humans to provide new conceptual frameworks [28,29]. Both MELVIN and THESEUS algorithms are available online [30,31]. These two algorithms take advantage of topological search, but there are also different refinements and proposals to combine multiple machine learning strategies. These include the use of neural networks [32] and reinforced learning [33], which also gives good results in the automated design of quantum communication protocols [34].

Compared to those approaches, our software does not use advanced machine learning techniques. We provide routines that can give exact numerical solutions and try to provide building blocks that can be later combined with these methods.

In that respect, the routines in QOptCraft focus on unitary matrices and analytic methods and they are more similar to the results from Gubarev and coworkers [35], where they search for heralded entangled states starting from random unitary transformations and a numerical optimization phase which is then refined analytically. Similarly, computers can be used to optimize the design of more general non-linear optical transformations which include losses, like in the quantum cloning experiments of [36]. Apart from those points of view, a polynomial description of the optical system also allows a different analytical design method using a Gröbner basis technique [37].

In most of this previous work, the methods have a desired target state and the state evolution is computed, like in QOptCraft, from the description of the classical elements. We use a slightly different, but equivalent, approach centered on the evolution unitary matrices and their corresponding Hamiltonians. Apart from that, QOptCraft also offers the ability to solve inverse problems directly from a full description of the desired evolution (Section 5.3).

For compatibility with other quantum optics design software, QOptCraft includes a series of functions that deal with states (Section 5.5.5). An example of their use for entangled state generation, which complements previous automated design methods, is given in Section 6.3.

## 3. The scattering matrix $S$ and the unitary evolution $U$

A linear interferometer acts on $m$ modes, or ports. The modes can be any collection of orthogonal states of a single photon. For simplicity we usually think of spatial modes: photons going through separate paths, but the results can be extended to modes representing orthogonal polarizations, different frequencies, separated time bins or photons with different values of orbital angular momentum.

Classically, for amplitudes $a_i$ in the $i$th mode, the corresponding output mode amplitude $b_i$ can be deduced from the scattering matrix $S$ of the system computing $\vec{b} = S\vec{a}$ [38]. The scattering matrix acting on the input amplitudes as

$$S\vec{a} = \begin{pmatrix} S_{11} & S_{12} & \ldots & S_{1m} \\ S_{21} & S_{22} & \ldots & S_{2m} \\ \vdots & \vdots & \ddots & \vdots \\ S_{m1} & S_{m2} & \ldots & S_{mm} \end{pmatrix} \begin{pmatrix} a_1 \\ a_2 \\ \vdots \\ a_m \end{pmatrix} \tag{1}$$

is unitary, which guarantees the conservation of energy in a passive linear network. The matrix description can be translated directly to the quantum evolution of a single photon where the field amplitudes are replaced by probability amplitudes and the matrix unitaricity guarantees that the probability of finding each state in the output superposition sums to 1 for unit vectors in the input, where the probabilities also sum to 1.

These unitary scattering matrices give the complete description of linear optical systems which preserve the number of photons. Losses and amplification can also be included using quasiunitary matrices [39] (Section 5.5.1).

The **quantum evolution** for $n$ photons is described by a unitary evolution matrix $U$ acting on all the possible photon states of $n$ photons distributed into $m$ modes. The resulting Hilbert space $\mathcal{H}_{m,n}$ has a dimension $M = \dim_{\mathbb{C}} \mathcal{H}_{m,n} = \binom{m+n-1}{n}$ which corresponds to counting all the possible ways to place $n$ "balls" (photons) into $m$ "boxes" (modes) (the multiset coefficient of $n$ and $m$ [40]).

We use a photon number notation. The state of the system is written as $|\psi\rangle = |n_1 n_2 \ldots n_m\rangle$, where $n_i$ is the number of photons found in the $i$th mode and $\sum_{i=1}^{m} n_i = n$.

In the QOptCraft package, we define a basis as an ordered collection of basis states in this format, which can be thought of as column vectors filled with zeroes and with a single 1 in the position corresponding to the index of the state.

For instance, for $n=3$ photons in $m=2$ modes, a basis would be $\{|30\rangle, |21\rangle, |12\rangle, |03\rangle\}$ and we can make the correspondence:

$$|30\rangle = \begin{pmatrix}1\\0\\0\\0\end{pmatrix}, \quad |21\rangle = \begin{pmatrix}0\\1\\0\\0\end{pmatrix}, \quad |12\rangle = \begin{pmatrix}0\\0\\1\\0\end{pmatrix}, \quad |03\rangle = \begin{pmatrix}0\\0\\0\\1\end{pmatrix}. \tag{2}$$

For that order, there is a unique $U$ giving the evolution of a linear interferometer with a scattering matrix $S$ (for which column $i$ gives the action on the $i$th input mode).

The functions in `QOptCraft` return the ordered list with the basis states so that all the objects are compatible.

Both $S$ and $U$ are unitary matrices and belong to the corresponding unitary groups: $S \in U(m)$, the group of unitary $m \times m$ matrices, and $U \in U(M)$, the group of unitary $M \times M$ matrices. Two matrices, $U_1, U_2$, which are equal up to a global phase, so that $U_1 = e^{i\Phi}U_2$, are equivalent from a physical point of view. There is no measurement that can tell them apart.

Notice that all the methods in the package could have been implemented for the corresponding matrices with determinant 1 in the special unitary groups $SU(m)$ and $SU(M)$. However, for simplicity and to keep the same framework as previous results, we have preferred to work with matrices from the unitary group instead of the special unitary group.

## 4. Structure of the problems: relation between the objects in a commutative diagram

The functions in `QOptCraft` allow the user to navigate through the commutative diagram in Fig. 1 which relates various matrices which appear when working with linear interferometers which preserve the photon number. The map $\varphi_{m,M}$ is a differentiable group homomorphism [9] and it induces an algebra homomorphism, $d\varphi_{m,M}$. In addition, $\varphi_{m,M}$ relates the unitary groups $U(m)$ and $U(M)$ containing the scattering matrix $S$ and the $n$-photon evolution operator $U$ and $d\varphi_{m,M}$ relates the algebras $\mathfrak{u}(m)$ and $\mathfrak{u}(M)$ whose elements correspond to antihermitian matrices $iH_S$ and $iH_U$ which give an equivalent description of the evolution through exponentiation of the Hamiltonians $H_S$ and $H_U$ ($S = e^{iH_S}$ and $U = e^{iH_U}$) [41,42].

$$\begin{array}{ccc} U(m) & \xrightarrow{\varphi_{m,M}} & U(M) \\ \exp\uparrow & & \uparrow\exp \\ \mathfrak{u}(m) & \xrightarrow{d\varphi_{m,M}} & \mathfrak{u}(M) \end{array}$$

**Fig. 1.** Commutative diagram describing the relationships between the different objects that define the evolution in linear optical systems.

## 5. Package overview

We will use the commutative diagram of Fig. 1 as a map to show how the different functions in the package work. Each family of functions is explained in terms of the theoretical results on which they are based. Many of the functions are implemented from independent methods for computing the same result, which can be used to check for consistency and to search for the most computationally efficient alternative for different input sizes.

### 5.1. The photonic homomorphism

The first function `StoU` returns the quantum unitary evolution $U$ for $n$ photons under the action of a linear interferometer specified by the user. The inputs are the number of photons, the scattering matrix of the system $S$ and an ordered basis of the state space so that the matrix $U$ is univocally determined.

The evolution is determined by the homomorphism $\varphi_{m,M}$ from the unitary group $U(m)$ to $U(M)$, in the upper part of the commutative diagram of Fig. 2 and it is computed with two different methods in the software.

$$\begin{array}{ccc} U(m) & \xrightarrow{\varphi_{m,M}} & U(M) \\ \exp\uparrow & & \uparrow\exp \\ \mathfrak{u}(m) & \xrightarrow{d\varphi_{m,M}} & \mathfrak{u}(M) \end{array}$$

**Fig. 2.** Evolution from $S$ (scattering matrix) to $U$ (quantum evolution) with the photonic homomorphism $\varphi_{m,M}$ (in red). (For interpretation of the colours in the figure(s), the reader is referred to the web version of this article.)

#### 5.1.1. Evolution from the Heisenberg picture

The matrix $U$ can be computed from the evolution of the creation operators that define all the possible input quantum states in the $\mathcal{H}_{m,n}$ Hilbert space. For an $n$-photon input state, with $\hat{a}_k^\dagger$ being the creation operation for mode $k$ [43],

$$|n_1 n_2 \ldots n_m\rangle = \prod_{k=1}^{m}\left(\frac{\hat{a}_k^{\dagger n_k}}{\sqrt{n_k!}}\right)|00\ldots 0\rangle. \tag{3}$$

An analysis in the Heisenberg picture [44–46] shows the quantum evolution can be expressed as:

$$U \, |n_1 n_2 \ldots n_m\rangle = \prod_{k=1}^{m} \frac{1}{\sqrt{n_k!}} \left( \sum_{j=1}^{m} S_{jk} \hat{a}_j^\dagger \right)^{n_k} |00 \ldots 0\rangle. \tag{4}$$

This method can be chosen in the function `StoU` with an optional argument.

*5.1.2. Computation using permanents*

There is also an alternative description related to the permanent of a matrix [45]. The permanent is a matrix function similar to the determinant which is computed without any position dependent sign correction. For an $n \times n$ matrix $A$, it can be defined as

$$\text{Per } A = \sum_{\sigma_n} \prod_{i=1}^{n} A_{i,\sigma(i)} \tag{5}$$

for column indices in all the permutations $\sigma$ in the symmetric group $S_n$.

The step from $S$ to $U$ depends on the number of photons. If we take the basis composed of the number states of Eq. (3), the element of $U$ that describes the transition from $|\Psi_{\text{in}}\rangle = |n_1\rangle_1 |n_2\rangle_2 \ldots |n_m\rangle_m$ to $|\Psi_{\text{out}}\rangle = |n'_1\rangle_1 |n'_2\rangle_2 \ldots |n'_m\rangle_m$ can be determined from $\langle n'_1|_1 \langle n'_2|_2 \ldots \langle n'_m|_m U |n_1\rangle_1 |n_2\rangle_2 \ldots |n_m\rangle_m$, which has a value

$$\frac{\text{Per}(S_{\text{in,out}})}{\sqrt{n'_1! \cdot n'_2! \cdots n'_m! \cdot n_1! \cdot n_2! \cdots n_m!}}. \tag{6}$$

In Eq. (6), $\text{Per}(S_{\text{in,out}})$ is the permanent of a matrix $S_{\text{in,out}}$ with elements $S_{i,j}$ from $S$ such that each row index $i$ appears exactly $n'_i$ times and each column index $j$ is repeated exactly $n_j$ times [45,9].

This method can be chosen in the function `StoU` in two different implementations. The first (for the argument `method = 1`) computes the permanents with a direct implementation and the second (for the argument `method = 2`) computes the permanent using Ryser's method [47], which tends to be faster for large state spaces in our experiments.

The direct implementation generates all the possible permutations of $n$ positions from Python's `itertools`.

In Ryser's algorithm the permanent is computed as a sum of products [48]. For an $n \times n$ matrix $A$, we use the formula

$$\text{Per } A = (-1)^n \sum_{X \subseteq \{1,2,\ldots,n\}} (-1)^{|X|} R(X) \tag{7}$$

where $X$ is any non-empty subset of the set of numbers from 1 to $n$; the formula sums the products

$$R(X) = \prod_{i=1}^{n} \sum_{j \in X} A_{i,j} \tag{8}$$

that multiply the sums of the column elements given by the indices in $X$ for each row. The signs avoid counting the same product twice.

*5.1.3. Efficiency*

None of the methods we have explained to compute $\varphi_{m,M}(S)$ is efficient. In fact, computing the permanent is known to be a PSPACE problem [49] and this, combined with other results, shows there is no known efficient classical method to compute the evolution in a general linear optical interferometer in the boson sampling problem [9]. Any classical method will face this complexity barrier. Nevertheless, the direct computation method in the package can be useful in the simulation of intermediate scale linear optical systems.

*5.2. Evolution of the effective Hamiltonian. The differential of the photonic homomorphism*

The matrix $H_U$ gives the effective Hamiltonian corresponding to the evolution $U$ through the exponential map with $U = e^{iH_U}$ [41]. The explicit Hermitian matrix can be deduced from the effective Hamiltonian $H_S$ of the linear multiport acting on one photon [42]. If we number the basis states, with $|q\rangle = |n_1 n_2 \ldots n_m\rangle$ and $|p\rangle = |n'_1 n'_2 \ldots n'_m\rangle$, the elements of $iH_U$ are

$$\langle p | iH_U | q \rangle = \langle p | \sum_{l=1}^{m} \sum_{j=1}^{m} iH_{S_{jl}} \hat{a}_j^\dagger \hat{a}_l | q \rangle. \tag{9}$$

The evolution corresponds to single photon processes from the weighted sum

$$iH_{U_{pq}} = \sum_{l=1}^{m} \sum_{j=1}^{m} i\sqrt{(n_j+1)n_l} H_{S_{jl}} \langle p | n_1 n_2 \ldots n_j + 1 \ldots n_l - 1 \ldots n_m \rangle. \tag{10}$$

The function `iHStoiHU` returns the Hermitian matrix $H_U$ corresponding to any given $H_S$ when the system is acting on $n$ photons. It is computed from the differential of the photonic homomorphism, $d\varphi_{m,M}$ (Fig. 3).

$$\begin{array}{ccc} U(m) & \xrightarrow{\varphi_{m,M}} & U(M) \\ \exp \uparrow & & \uparrow \exp \\ \mathfrak{u}(m) & \xrightarrow[d\varphi_{m,M}]{} & \mathfrak{u}(M) \end{array}$$

**Fig. 3.** Evolution from $H_S$ (the single photon Hamiltonian) to $H_U$ (the n-photon Hamiltonian) with the differential of the photonic homomorphism $d\varphi_{m,M}$ (in red).

### 5.3. Inverse problems. Design of linear interferometers.

In many cases we know the target evolution $U$ we would like to obtain and we need to find out the experimental setup that produces that evolution.

In general, linear optics offers only a limited range of all the possible evolutions in the Hilbert space $\mathcal{H}_{m,n}$ for $n$ photons. For more than one photon or one port, it is clear that the degrees of freedom in the scattering matrix which gives the experimental implementation are less than the required degrees of freedom to produce any $U \in U(M)$ [50–52].

The image im$\varphi$ of $\varphi_{m,M}$ is a subgroup of $U(M)$ and, as the number of photons grows, the evolutions $U$ which can be actually realized with linear optics become a smaller set of all the possible unitary evolutions.

The package contains two different methods for design, depending on whether the desired evolution is possible or must be approximated

The function `SfromU` serves as an initial check. When given a target evolution $U$ for $n$ photons in $m$ modes, it answers whether there exists a valid linear evolution or not. If it is possible, the function returns the matrix $S$ which provides the desired evolution.

The inverse homomorphism $\varphi_{m,M}^{-1}$ in Fig. 4 is computed with the help of the adjoint representation of $U$ using the projections over the basis of the $\mathfrak{u}(m)$ and $\mathfrak{u}(M)$ algebras [53].

$$\begin{array}{ccc} U(m) & \xrightleftharpoons[\varphi_{m,M}^{-1}]{\varphi_{m,M}} & U(M) \\ \exp \uparrow & & \uparrow \exp \\ \mathfrak{u}(m) & \xrightarrow[d\varphi_{m,M}]{} & \mathfrak{u}(M) \end{array}$$

**Fig. 4.** The scattering matrix $S$ can be recovered from the unitary evolution of the $n$ photons, $U$, with the inverse of the photonic homomorphism $\varphi_{m,M}^{-1}$ (in red).

Internally, it generates a basis for the image algebra giving the effective Hamiltonian $H_U \in \mathfrak{u}(M)$ which corresponds to the evolution through the differential $d\varphi_{m,M}$ (`iHStoiHU`) of the canonical basis of $\mathfrak{u}(m)$.

The adjoint representation gives a way to compute the $H_U$ corresponding to $U$. While the matrix exponential gives an efficient way to go from $H_S$ to $S$ or from $H_U$ to $U$, matrix logarithms are not trivial for this application (see Section 5.5.2). In particular, we need to restrict to logarithms in the image subalgebra of $\mathfrak{u}(m)$, which is smaller than $\mathfrak{u}(M)$.

The basis of the image algebra allows us to write the inverse problem of going from $U$ to $S$ as a linear system of equations. If the system is incompatible, we declare $U$ impossible to produce with linear optics alone. Otherwise, solving the system gives an $S$ such that $U = \varphi(S)$.

When the system has no solution and $U$ cannot be exactly implemented with linear optics, there is a second function, `Toponogov`, which finds the scattering matrix $S$ providing a quantum evolution $\tilde{U} = \varphi(S)$ that approximates $U$.

The approximation is designed to minimize $\left\| U - \tilde{U} \right\|$ for the Frobenius norm of the difference between the desired evolution $U$ and the evolution $\tilde{U}$ what we can provide with linear optics. The function `Toponogov` works by an iterative procedure based on Toponogov's comparison theorem [54] using an initial guess matrix $U_0$ inside the image of $\varphi_{m,M}$. The results are locally optimal insofar the method finds the matrix $\tilde{U}$ which can be implemented which is closest to $U$ in the local neighbourhood of $U_0$. This method can help to refine a random search. At the bare minimum it will return the original random guess, usually with an improved approximation after just a few steps.

The initial guesses are generated by the function `RandU` which produces a random matrix in the image group, $U_r \in \text{im}\varphi$ (see Section 5.5.3). The function first generates a random unitary $S_r$ chosen uniformly at random from $U(m)$ and, then, it computes the evolution for $n$ photons with `StoU` to produce $U_r = \varphi(S_r)$.

With different random initial guesses, we can explore the solution landscape and avoid staying trapped in a single local minimum.

### 5.4. Experimental realizations of linear interferometers

The scattering matrix $S$ gives a complete description of a linear interferometer and can be computed for any experimental setup by taking the product of the scattering matrices of the corresponding elements.

There are two basic optical elements: beam splitters and phase shifters. Beam splitters acting on two modes have a scattering matrix

$$S_{BS} = \begin{pmatrix} \cos\omega & \sin\omega \\ \sin\omega & -\cos\omega \end{pmatrix}. \tag{11}$$

Phase shifters introduce a relative phase $\Phi$ in one mode with respect to the rest and have a scattering matrix

$$S_{PS} = \begin{pmatrix} 1 & 0 \\ 0 & e^{i\Phi} \end{pmatrix} \tag{12}$$

for two modes. For larger systems with $m$ modes, the resulting scattering matrix is an $m \times m$ identity matrix where the elements of $S_{BS}$ or $S_{PS}$ replace the elements in the rows and columns with indices $(k,k)$, $(k,l)$, $(l,k)$ and $(l,l)$ when acting on the modes with index $k$ and $l$.

Any sequence of $L$ elements, with $S_1$ the first element at the input and $S_L$ the last one, is described by the product matrix $S = S_L S_{L-1} \cdots S_1$.

There are a few known constructive methods which, given any valid unitary $S$, produce a list of the beam splitters and phase shifters required to give that $S$ [55–57].

The package includes the function Selements which, for any input unitary matrix $S$, returns the list of the required elements for experimental implementation. The user can choose from the decompositions of Clements, Humphreys, Metcalf, Kolthammer and Walsmley [56] (default) or the Reck, Zeilinger, Bernstein and Bertani [55] method.

The output of the function Selements can be used as a list of the elements needed to obtain the scattering matrix $S$ in an optical table or to choose the parameters of programmable integrated photonic chips [58]. Combined with the design methods in Section 5.3, it gives a complete design path from the desired unitary to an experimental realization.

### 5.5. Other methods

The package is completed with some useful intermediate functions and partial generalizations that help to perform numerical experiments for research.

#### 5.5.1. Lossy linear interferometers and squeezing

The decompositions described in Section 5.4 are given for ideal, lossless multiports. The description of linear optical systems in terms of a scattering matrix can be generalized to introduce losses and squeezing if we replace unitary matrices by what can be described as quasiunitary matrices $S$, such that

$$SGS^\dagger = G \qquad (13)$$

where $G$ is a diagonal block matrix with elements $I$ and $-I$ for $I$ the $m \times m$ identity matrix in a system with $m$ modes.

In this framework, we need to work both with creation and destruction operations and the size of $S$ doubles [41]. Now

$$S\vec{a} = \begin{pmatrix} A & B \\ B^* & A^* \end{pmatrix} \begin{pmatrix} \vec{a} \\ \vec{a}^\dagger \end{pmatrix}, \qquad (14)$$

where $\vec{a}$ and $\vec{a}^\dagger$ are vectors with the annihilation and creation operators for each mode and the matrix blocks at the bottom are the (untransposed) complex conjugates of the first row blocks. For the passive networks discussed in the rest of the paper $B = 0$.

The package includes the function QuasiU giving a decomposition of any quasiunitary matrix in terms of simple optical blocks, which now include amplification and losses. The function follows the decomposition of Tischler, Rockstuhl and Słowik [39].

This method uses a different underlying description of the optical system and it works independently from the other functions. The rest of the fuctions involving matrix operations included in the package assume the inputs are unitary matrices. Thus, the only scenario these quasiunitary samples would be compatible with them would be for $B = 0$, where there is no need for two $A$, $A^*$ blocks due to no crossing between loss and gain. If $B = 0$, we can simply use $A$ as the unitary input to any other method in the package.

#### 5.5.2. Matrix logarithms

One particular difficulty when going from the unitary groups to the their corresponding unitary algebras is choosing the right branch of the matrix logarithm.

While the exponential map is injective and there is only one unitary evolution corresponding to each effective Hamiltonian, the logarithm is a multivalued function and there exist many possible Hamiltonians for the same evolution.

The package works with the **principal branch of the logarithm**, $\log U = iK$, of a unitary matrix $U \in U(m)$, defined as

$$K^\dagger = K, \quad \exp(iK) = U, \quad \text{with the eigenvalues of } K \text{ in } (-\pi, \pi]. \qquad (15)$$

There are two slightly different definitions for the principal matrix logarithm. We choose this definition with a closed interval on one side instead of the definition with a valid interval $(-\pi, \pi)$ for the eigenvalues so that there is always a principal matrix logarithm, even for matrices with real negative eigenvalues $(-1)$. Under our definition, the matrix logarithm loses some properties, like infinite differentiability, but they are not needed in any calculation involving linear interferometers. Covering all the possible input matrices is more important for consistency and the $-1$ eigenvalue appears in many interesting evolutions, like the Quantum Fourier Transform (described in Section 5.5.4).

There are efficient and stable numerical methods that can compute the principal logarithm of a unitary matrix. The function Matlog uses different iterative algorithms from Loring's paper [59] under the user's choice.

There is one final warning. One might be tempted to use the matrix logarithm of a matrix $U$ to compute $H_U$ and then use the basis of the image algebra, available with the function AlgBasis, to find a suitable implementation. Unfortunately, the principal logarithm is not guaranteed to lie inside the unitary image algebra, which is a subalgebra of $\mathfrak{u}(M)$. The correct method is given in the function SfromU, which solves the problem by taking an indirect route in the adjoint representation.

The package can be used to perform numerical experiments that show the branch of the logarithm inside the image algebra is not predictable in a straightforward manner.

### 5.5.3. Random unitaries

Generating unitaries uniformly at random in $U(m)$ and in $\text{im}\varphi \subset U(M)$ can be interesting in numerical experiments where we want to sample random interferometers and when testing new functions. It is also a fundamental element in the approximation method in the function `Toponogov` described in Section 5.3.

The function `RandU` generates a unitary matrix chosen uniformly with respect to the Haar measure. The random unitary is generated from a random complex matrix with elements with normal random real and imaginary parts which is transformed by a QR decomposition and then normalized [60,61].

The function `RandImU` generates a random unitary $U \in \text{im}\varphi \subset U(M)$ with a random $m \times m$ matrix $S_R$ generated with `RandU`, which is then taken into $U(M)$ using `StoU`. The results is a random matrix in the image.

`RandU` is valuable for testing. For instance, we have used it to sample randomly from $U(M)$ and then check the approximations produced by `Toponogov`.

### 5.5.4. Quantum Fourier transform matrices

The package includes the function `QFT`to generate a Quantum Fourier Transform of any chosen size. This matrix has many interesting properties and is at the heart of many successful quantum algorithms, like Shor's algorithm for integer factorization [62].

An $N \times N$ QFT matrix has elements

$$QFT_{x,y} = \frac{1}{\sqrt{N}} e^{i\frac{2\pi xy}{N}} \tag{16}$$

for $x, y = 0, \ldots, N-1$ in row $x+1$ and column $y+1$.

The QFT unitary is simply a normalized Discrete Fourier Transform matrix, which is an entangling operation and has multiply degenerate eigenvalues (which can only take the values $1, -1, i, -i$) [63,64]. These properties make it a good "hard case" that can serve as benchmark for our numerical methods. Additionally, it is the scattering matrix describing symmetric optical networks [65,66].

### 5.5.5. State routines and entanglement evaluation

In many applications, individual quantum states are more important than the whole evolution matrix. `QOptCraft` also includes some functions that make working with states easier.

The function `leading_terms` counts the most relevant elements of a superposition and `state_leading_fidelity` returns a clean state taking only the terms of the superposition that contribute up to a certain fidelity $F$, i.e. the truncated state $\left|\tilde{\psi}\right\rangle$ is close to the original one $|\psi\rangle$ so that $|\langle\psi|\tilde{\psi}\rangle|^2 > F$. The function `state_leading_terms` returns an approximation that rounds to zero the probability amplitude of the terms of the superposition that have a probability of being measured below a given threshold.

The vector representation in the whole state space can be a bit cumbersome. `state_in_basis` takes a natural description of a state given as a list of the terms of the superposition and their complex weights and returns a complex array in the corresponding basis in the format `QOptCraft` uses internally.

Finally, in some occasions we need to evaluate whether a state presents entanglement or not. There is no single entanglement measure which captures the whole phenomenon and quantifies it in a clear cut way [67], but they are a useful guide. We will use as an orientation the Schmidt rank, which assigns a figure to the entanglement in bipartite systems [68].

The function `schmidt_rank_vector` evaluates the entanglement between different subsystems of the global state. It returns a vector where each element is the Schmidt rank for the bipartite system composed of the corresponding subsystem and the rest of the state. The user can introduce different groupings of the modes to define each subsystem. Internally, it is computed by taking the state space to a larger dimension where all the modes can carry up to $n$ photons.

A detailed example of its use for entanglement evaluation can be found in Section 6.3.

### 5.5.6. Applications to quantum information

Apart from experiments in quantum optics, we might be interested in using a linear interferometer as the physical implementation for a target unitary quantum gate $U_t$ which performs a useful operation in quantum information processing. This can be part of a quantum algorithm or a transformation in a quantum communication protocol.

However, in linear interferometers, certain questions like swapping between two states $|n_1 \ldots n_m\rangle$ and $|n'_1 \ldots n'_m\rangle$ are not possible in all the cases. For instance, no linear operation can take $|20\rangle$ into $|11\rangle$ deterministically.[1]

For that reason, if $U_t$ is a logical quantum operation, the mapping between the physical and the logical states is not trivial. The function `SfromU` includes a parameter perm, enabling (for perm = *True*) its application for all available ordering of the basis states in $\mathcal{H}_{m,n}$, to make sure a logical operator is not possible or give the scattering matrix which allows its experimental realization.

## 6. Usage examples

### 6.1. Computing the unitary evolution U from the scattering matrix S. Comparison of the methods

In most applications we will need to compute the evolution for a known linear optical system. `QOptCraft` offers four alternative methods to compute $U = \varphi(S)$ which can be specified when calling `StoU`. Two of them are based on the state transformations given from the description of the evolution of the operators in the Heisenberg picture shown in Equation (4), method=0, or the description in terms of permanents [45] computed either from the direct definition of the permanent, method=1 or with an implementation using Ryser's algorithm [47,48], method=2.

---

[1] Using Equation (4) we can see $U|20\rangle = \frac{1}{\sqrt{2}}(S_{11}^2 \hat{a}_1^{\dagger 2} + 2S_{11}S_{21}\hat{a}_1^\dagger \hat{a}_2^\dagger + S_{21}^2 \hat{a}_2^{\dagger 2})|00\rangle$. The output state can only be $|11\rangle$ for a linear interferometer where both $S_{11}S_{21} = \frac{1}{\sqrt{2}}$ and $S_{11} = S_{21} = 0$ are simultaneously true, which is impossible.

**Table 1**
Logarithm of the normalized execution times (arbitrary units) for the four computation methods and different sizes $M$ of the relevant Hilbert space.

| $m$ | 2 | 2 | 2 | 2 | 3 | 3 | 3 | 3 | 4 | 4 | 4 | 4 | 5 | 5 | 5 | 5 |
|---|---|---|---|---|---|---|---|---|---|---|---|---|---|---|---|---|
| $n$ | 2 | 3 | 4 | 5 | 2 | 3 | 4 | 5 | 2 | 3 | 4 | 5 | 2 | 3 | 4 | 5 |
| $M$ | 3 | 4 | 5 | 6 | 6 | 10 | 15 | 21 | 10 | 20 | 35 | 56 | 15 | 35 | 70 | 126 |
| Heisenberg (0) | 0.00 | 1.89 | 3.54 | 3.80 | 2.84 | 4.85 | 5.44 | 6.54 | 4.11 | 5.44 | 7.29 | 9.31 | 5.03 | 6.73 | 9.22 | 11.75 |
| Permanent (1) | 0.31 | 1.81 | 3.57 | 4.50 | 3.58 | 2.47 | 3.92 | 5.98 | 3.51 | 3.84 | 5.59 | 7.91 | 3.10 | 4.98 | 7.00 | 9.55 |
| Ryser (2) | 0.34 | 2.23 | 3.75 | 3.03 | 3.85 | 2.92 | 4.22 | 5.52 | 3.85 | 4.29 | 5.91 | 7.47 | 3.41 | 5.43 | 7.30 | 9.10 |
| Hamiltonian (3) | 1.49 | 2.78 | 2.74 | 2.82 | 3.89 | 4.51 | 5.19 | 5.67 | 5.18 | 6.10 | 7.38 | 8.65 | 5.90 | 7.88 | 9.82 | 11.52 |

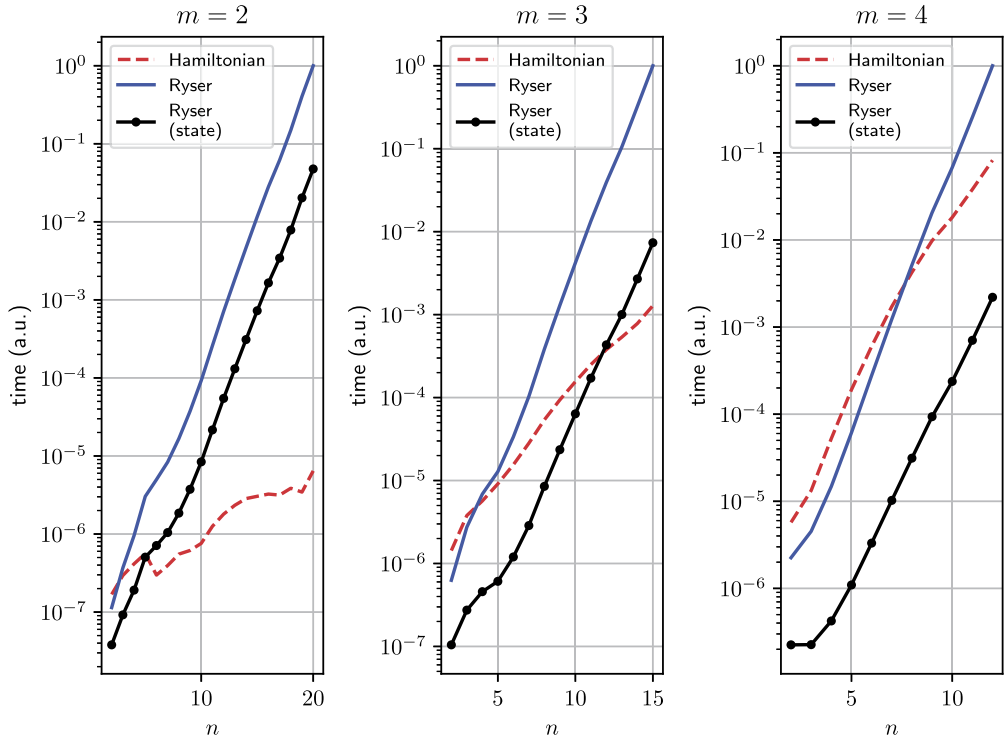

**Fig. 5.** Execution time comparison for Ryser's algorithm (solid line) and the Hamiltonian methods (dotted line) with an increasing number of photons ($x$ axis). The times are shown in arbitrary units, normalized to the maximum time and presented in logarithmic scale. The line with the point markers represents the time for Ryser's method divided by the number of columns $M$ of the matrix $U$ and gives an approximation to the time needed to compute the evolution of a single state.

Internally, these methods build the unitary matrix $U$ column by column by computing how each input state in the state basis evolves in a linear system with the given scattering matrix $S$.

For individual state evolution, the functions evolution, evolution_2 and evolution_2_ryser can be called on individual states of the basis to recover the corresponding output state without computing the full matrix using methods 0, 1 and 2, respectively.

The fourth method to compute $U = \varphi(S)$, method=3, is indirect. First we take the matrix logarithm of $S$ and evolve this Hamiltonian for input states with $n$ photons following Equation (9). Then, $U$ is computed from the matrix exponential of this Hamiltonian $H_U$.

The function QOCTest runs tests to compare the efficiency of these methods under different circumstances and can take the output to an Excel table. Table 1 compares the time it takes the four different methods to compute $U = \varphi(S)$ for a scattering matrix $S$ chosen uniformly at random using RandU. QOCTest can also compare the times starting from $S$ matrices that are the $m \times m$ QFT (a symmetric multiport) with similar results.

The results in the Table are presented in arbitrary time units scaled by the minimum computation time and shown in logarithmic scale for a better comparison. These tests were performed on an AMD Phenom II X6 1055T Processor with 6 cores of 2.8 GHz where the execution is confined to a single core.

By default, QOptCraft uses the permanent method with Ryser's algorithm. However, it is interesting to note that, for a fixed number of modes, we have observed the Hamiltonian method tends to give a better performance as the number of photons grow.

Fig. 5 shows the results of three experiments where, for a random scattering matrix $S_r$ chosen uniformly at random from the Haar measure, we computed $U = \varphi(S_r)$ using Ryser's algorithm and the Hamiltonian method. The experiments were performed in a single 2.8 GHz core for $m = 2$, $m = 3$ and $m = 4$ and a growing number of photons $n$ (up to the point where the execution times were reasonable). The results are presented in logarithmic scale and the time units are normalized by the maximum execution time. All the times are computed from the average of 5 executions for the same matrix.

We have added for comparison a line with the time for Ryser's method divided by the dimension $M$ of the state space. This approximates the time to compute a single column of the matrix and gives an idea of the work involved in computing the evolution of a single state as opposed to computing the full matrix. For those tasks, it might be better to use one of the state by state methods instead of computing the whole matrix $U$, but, at least in our examples, there is a threshold at which it is more efficient just to compute the whole matrix $U$ from its Hamiltonian.

Notice that, in this work, there has been no explicit optimization to reduce the running time apart from basic tweaks. It would interesting to see in a future work whether the matrix exponential can be sped up taking advantage that $H_U$ is a sparse matrix and how the Hamiltonian method compares to optimized versions of Ryser's algorithm.

*6.2. The Quantum Fourier Transform evolution*

As an example on the package, we show the design of a Quantum Fourier Transform evolution for a system with 3 ports and 2 photons ($n = 2$, $m = 3$). The state space has a size $M = \binom{n+m-1}{n} = \binom{4}{2} = 6$.

In our design example, we first generate the target $6 \times 6$ unitary matrix

$$U_t = \frac{1}{\sqrt{6}} \begin{pmatrix} 1 & 1 & 1 & 1 & 1 & 1 \\ 1 & e^{i\frac{2\pi}{6}} & e^{i\frac{2\pi 2}{6}} & e^{i\frac{2\pi 3}{6}} & e^{i\frac{2\pi 4}{6}} & e^{i\frac{2\pi 5}{6}} \\ 1 & e^{i\frac{2\pi 2}{6}} & e^{i\frac{2\pi 4}{6}} & e^{i\frac{2\pi 6}{6}} & e^{i\frac{2\pi 8}{6}} & e^{i\frac{2\pi 10}{6}} \\ 1 & e^{i\frac{2\pi 3}{6}} & e^{i\frac{2\pi 6}{6}} & e^{i\frac{2\pi 9}{6}} & e^{i\frac{2\pi 12}{6}} & e^{i\frac{2\pi 15}{6}} \\ 1 & e^{i\frac{2\pi 4}{6}} & e^{i\frac{2\pi 8}{6}} & e^{i\frac{2\pi 12}{6}} & e^{i\frac{2\pi 16}{6}} & e^{i\frac{2\pi 20}{6}} \\ 1 & e^{i\frac{2\pi 5}{6}} & e^{i\frac{2\pi 10}{6}} & e^{i\frac{2\pi 15}{6}} & e^{i\frac{2\pi 20}{6}} & e^{i\frac{2\pi 25}{6}} \end{pmatrix}. \quad (17)$$

The function `QFT` generates the matrix in the output file `QFT_matrix_6.txt`.

```
1    from QOptCraft.Main_Code import *  # load all functions
2    # We first generate the 6 x 6 QFT matrix:
3    QFT(filename="QFT_matrix_6",N=6)
```

Our first step is to find whether the matrix in Eq. (17) can be obtained exactly with linear optics or not, for which we use the function `SfromU`.

```
4    # Is the original matrix already possible?
5    SfromU(file_input=True,filename="QFT_matrix_6",
6    txt=True,acc_d=2,m=3,n=2)
```

One of the possible input parameters is an ordered basis for the Hilbert space with $n = 2$ photons in $m = 3$ modes. In this case, we have left it empty and `SfromU` generates the default basis, which, in this case, is $\{|200\rangle, |110\rangle, |101\rangle, |020\rangle, |011\rangle, |002\rangle\}$. We have

$$|200\rangle = \begin{pmatrix}1\\0\\0\\0\\0\\0\end{pmatrix}, \quad |110\rangle = \begin{pmatrix}0\\1\\0\\0\\0\\0\end{pmatrix}, \quad |101\rangle = \begin{pmatrix}0\\0\\1\\0\\0\\0\end{pmatrix}, \quad |020\rangle = \begin{pmatrix}0\\0\\0\\1\\0\\0\end{pmatrix}, \quad |011\rangle = \begin{pmatrix}0\\0\\0\\0\\1\\0\end{pmatrix}, \quad |002\rangle = \begin{pmatrix}0\\0\\0\\0\\0\\1\end{pmatrix}. \quad (18)$$

The code is **incapable** of finding a valid $S$ which produces the quantum evolution $U$ of the QFT matrix of Eq. (17). This was bound to be expected, since linear optics elements can only produce a limited amount of evolutions $U \in U(M)$.

If we want to continue, we need to turn to the iterative algorithm in `Toponogov` (for $m = 3$ modes, $n = 2$ photons and the matrix file with the $6 \times 6$ QFT).

```
7    # Getting "QFT_matrix_6.txt"'s closest evolution matrix U.
8    Toponogov(file_input=True,filename="QFT_matrix_6",base_input=False,
     ↪ file_output=True,m=3,n=2,tries=20)
```

After 20 attempts, we found 9 different $U_t^i$ approximations, with some approximations appearing more than once. Each approximation has a different distance to the original $U_t$ in (17). We use as a distance the Frobenius norm of $U_t - U_t^i$ defined by our matrix inner product metric. `Toponogov` produces a general output file listing all the found evolution matrices and their distance to the target evolution, as well as files with each individual matrix.

In our randomized trial, we choose the third one $U_t^3$, with a minimal distance $d(U_t^3, U_t) = 2.29449$, and elements:

$$U_t^3 = \begin{pmatrix} 0.01i & -0.08-0.05i & -0.03-0.05i & 0.71-0.11i & 0.62+0.11i & 0.24+0.14i \\ 0.03-0.02i & -0.07+0.26i & -0.16+0.19i & 0.12+0.61i & 0.06-0.37i & 0.27-0.51i \\ 0.09+0.05i & -0.77+0.18i & -0.46-0.04i & 0.14-0.09i & -0.20+0.10i & -0.26+0.02i \\ -0.10-0.09i & -0.26-0.07i & 0.33+0.23i & -0.26+0.06i & 0.56+0.07i & -0.54-0.27i \\ 0.10-0.47i & -0.12-0.41i & -0.05+0.59i & 0.06+0.08i & -0.10-0.27i & 0.03+0.38i \\ 0.82-0.26i & -0.14+0.11i & 0.41-0.21i & 0.01-0.02i & -0.05+0.05i & 0.10-0.07i \end{pmatrix}. \tag{19}$$

We are guaranteed to find a suitable $S_t$ matrix which produces the approximate QFT matrix in Eq. (19) for the evolution of 2 photons in 3 modes and the basis in Eq. (18). Using Selements, we can decompose $S_t$ into different linear optic devices.

```
9    # Getting "QFT_matrix_6_toponogov_3.txt"'s S-matrix.
10   SfromU(file_input=True,filename="QFT_matrix_6_toponogov_3",
11   file_output=True,m=3,n=2)
12
13   # Decomposition of "QFT_matrix_6_toponogov_3.txt's S-matrix".
14   Selements(file_input=True,file_output=True,newfile=False,
15   impl=0,filename="QFT_matrix_6_toponogov_3_m_3_n_2_S_recon_main")
```

SfromU takes the file with the best approximation to the QFT and generates a file with the scattering matrix $S_t$ which gives the desired approximated evolution. The output is then used as the input of Selements to obtain a list of the basic optical devices needed to build $S_t$ experimentally.

The resulting $S_t$ is:

$$S_t = \begin{pmatrix} 0.07679 & -0.61787+0.57579i & -0.48484+0.21387i \\ -0.11099-0.34803i & -0.36813-0.36367i & 0.32869+0.70053i \\ 0.63057-0.68047i & -0.05348+0.12676i & 0.19068-0.28992i \end{pmatrix}. \tag{20}$$

By default, Selements gives a list of phase shifters in a diagonal matrix $D$ and a list of $T_{mn}$ matrices, following the decomposition in [56] so that:

$$S_t = (T_{12}^{even})^{-1} (T_{23}^{even})^{-1} D T_{12}^{odd}. \tag{21}$$

After running Selements, we obtain the matrices:

$$D = \begin{pmatrix} 0.7024+0.7118i & 0.0000 & 0.0000 \\ 0.0000 & -0.3887+0.9214i & 0.0000 \\ 0.0000 & 0.0000 & 0.5495-0.8355i \end{pmatrix}, \tag{22}$$

$$T_{12}^{odd}(\theta = 1.7180, \phi = 0.3481) = \begin{pmatrix} -0.1379-0.0500i & -0.9892 & 0.0000 \\ 0.9298+0.3374i & -0.1467 & 0.0000 \\ 0.0000 & 0.0000 & 1.0000 \end{pmatrix}, \tag{23}$$

$$T_{12}^{even}(\theta = -2.5368, \phi = -1.5941) = \begin{pmatrix} 0.0192+0.8249i & 0.5650 & 0.0000 \\ 0.0131+0.5649i & -0.8251 & 0.0000 \\ 0.0000 & 0.0000 & 1.0000 \end{pmatrix}, \tag{24}$$

$$T_{23}^{even}(\theta = 1.2164, \phi = 1.0205) = \begin{pmatrix} 1.0000 & 0.0000 & 0.0000 \\ 0.0000 & 0.1815+0.2958i & -0.9379 \\ 0.0000 & 0.4905+0.7994i & 0.3470 \end{pmatrix}. \tag{25}$$

The output is given in the format $T_{k,l}$ for a generalized beam splitter acting on modes $k$ and $l$. The resulting optical element is defined by two angles $\phi$ and $\theta$. $\phi$ gives the phase shift in a phase shifter at the input $k$ and $\theta$ gives the splitting ratio of a general beam splitter so that

$$T_{k,l}(\theta, \phi) = \begin{pmatrix} e^{i\phi}\cos\theta & -\sin\theta \\ e^{i\phi}\sin\theta & \cos\theta \end{pmatrix}. \tag{26}$$

$D$ is achieved with $m$ phase shifters.

Each $T_{k,l}$ evolution can be achieved with a phase shifter with a $\phi$ phase shift followed by a beam splitter with a splitting angle $\theta$. The inverse operations $(T_{k,l})^{-1}$ can be achieved with a beam splitter with a splitting angle $-\theta$ followed by a phase shifter with a $-\phi$ shift.

This completes the whole path from the target evolution in Eq. (17) to the experimental setup which best approximates that evolution for two input photons in the ordered basis of Eq. (18). Fig. 6 shows the final optical setup corresponding to the results in the output of Selements.

Alternatively, the same evolution can be achieved replacing the unbalanced beam splitter by two balanced, $50:50$, beam splitters with a phase shift $2\theta$ in the middle and the phase shifters in $D$ can be avoided if the output is measured immediately after the device with some transformations on the $T_{k,l}$ matrices [56].

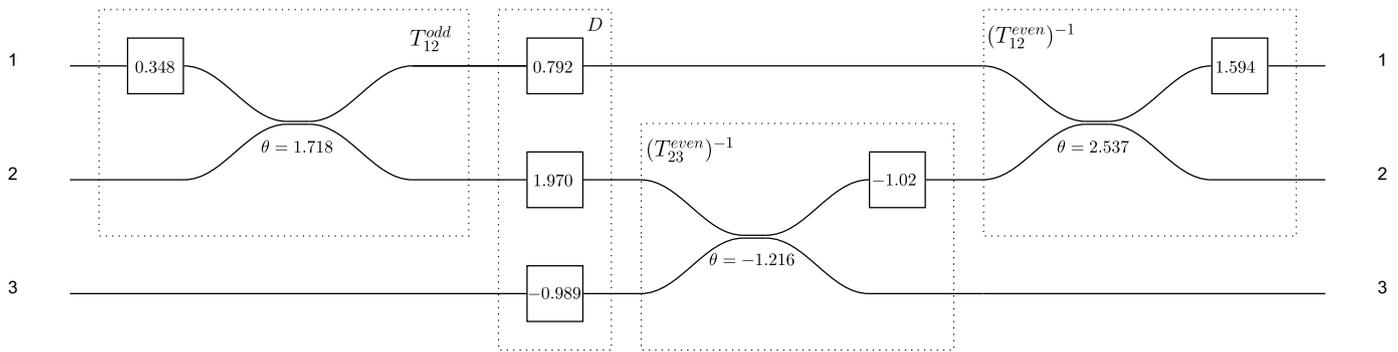

**Fig. 6.** Optical system giving the optimal approximation to a 6 × 6 QFT for 2 photons using unbalanced optical couplers (or beam splitters) and phase shifters (phase indicated inside the box).

*6.3. Generation of entangled states*

*6.3.1. Quantifying entanglement*

Highly entangled states are a basic resource in multiple quantum optics experiments and quantum information procotols. For instance, the Bell state

$$\frac{|00\rangle + |11\rangle}{\sqrt{2}}, \tag{27}$$

which can be generated with nonlinear crystals, is usually employed as an input to linear optical systems to provide additional capacities to the system.

Generating these entangled resource states can be complex and there exist different proposals to create a variety of these states as efficiently as possible [21,35]. One of the applications of `QOptCraft` is producing and evaluating optical transformations which produce highly entangled photon states from easy to produce inputs.

The function `schmidt_rank_vector` returns an orientiative score on how entangled the different subsystems in a state are. A value of 1 means the subsystems are separable. While, as a rule, higher numbers in all the positions mean a higher available entanglement, the Schmidt rank vector must be used with care.

We can check some basic examples. For instance, in our mode notation, a Bell state with polarization correlation for photons in a horizontal $|H\rangle$ or vertical $|V\rangle$ polarization state can be represented as

$$\frac{|0101\rangle + |1010\rangle}{\sqrt{2}} \tag{28}$$

with $|HH\rangle \equiv |0101\rangle$ and $|VV\rangle \equiv |1010\rangle$.

```
1  basisn2m4=photon_combs_generator(4,[1,0,1,0])
2  Bell=state_in_basis([[1,0,1,0],[0,1,0,1]],
       [1/np.sqrt(2),1/np.sqrt(2)],basisn2m4)
3  print('schmidt_rank_vector(Bell,basisn2m4,[2,2]))
```

The resulting vector [2,2] suggests it is an entangled state for the two subsystems with two modes each.

The Schmidt rank vector is useful to detect entanglement, but should be used with care. For instance, the state $\sqrt{1-\varepsilon}\,|0101\rangle + \sqrt{\varepsilon}\,|1010\rangle$ for $\varepsilon \ll 1$ is basically a separable state $|0101\rangle$. However, the code

```
4  eps=1e-3
5  AlmostSeparable=state_in_basis([[1,0,1,0],[0,1,0,1]],
       [np.sqrt(1-eps),np.sqrt(eps)],basisn2m4)
6  print(schmidt_rank_vector(AlmostSeparable,basisn2m4,[2,2]))
```

returns the same result as the proper Bell state. The state handling functions can help to clean up the input states. For a threshold fidelity of 0.99, the code

```
 7   fidelity=0.99
 8   states,weights=state_leading_fidelity(AlmostSeparable,
       ↪ basisn2m4,fidelity)
 9   cleanstate=state_in_basis(states,weights,basisn2m4)
10   print(schmidt_rank_vector(cleanstate,basisn2m4,[2,2]))
```

returns $[1, 1]$ showing the input was, in essence, a separable state.

Finally, we can use `schmidt_rank_vector` to study higher order entanglement. We can see an application to photons carrying Orbital Angular Momentum, OAM, with the state $|\psi_{422}\rangle$ that was found using automated search in [21]. We have a state

$$|\psi_{422}\rangle = \frac{1}{2}(|000\rangle + |101\rangle + |210\rangle + |311\rangle) \tag{29}$$

where each ket represents the state of a single photon and the numbers give the mode, which corresponds to the OAM value the photon has. The first photon can carry OAM values from 0 to 3 and last two photons can only be in states with OAM 0 or 1. In the mode notation of `QOptCraft`, the state becomes

$$|\psi_{422}\rangle = \frac{1}{2}(|0001\rangle |01\rangle |01\rangle + |0010\rangle |01\rangle |10\rangle + |0100\rangle |10\rangle |01\rangle + |1000\rangle |10\rangle |10\rangle) \tag{30}$$

where the first four modes give the possible state of the first photon, the following two modes the state of the second photon and the last two modes the state of the third photon.

If we generate this state in the corresponding basis and check the Schmidt rank for the subsystems of each photon with the rest with the commands

```
11   basisn3m8=photon_combs_generator(8,[1,1,1,0,0,0,0,0])
12   State422=state_in_basis([[0,0,0,1,0,1,0,1],
       ↪ [0,0,1,0,0,1,1,0],[0,1,0,0,1,0,0,1],[1,0,0,0,1,0,1,0]],
       ↪ [1/2,1/2,1/2,1/2],basisn3m8)
13   print(schmidt_rank_vector(State422,basisn3m8,[4,2,2]))
```

we obtain, as expected, the $[4, 2, 2]$ vector.

*6.3.2. Creation of advanced entangled states*

In practice, we would like to be able to generate highly entangled states from inputs that can be efficiently generated in the lab. This usually means finding an optical system which can take a simple separable input with a single photon in a few selected modes and produce a useful entangled state which can be used in other protocols.

The output states should be easy to recognize and have a simple description.

A random $S$ matrix with a separable input produces output states that are highly entangled. We can check this is the case using

```
1   # Entanglement produced by a random transformation
2   basisn4m5=photon_combs_generator(5,[1,0,1,1,1])
3   Sr=RandU(file_output=False,filename=False,N=5,txt=False)      # The
      ↪ random scattering matrix
4   Ur,bas=StoU(file_input=False,S=Sr,file_output=False,
      ↪ filename=False,method=2,n=4,vec_base=basisn4m5) # Evolution U
5   input_state=state_in_basis([[1,1,1,1,0]],np.array([1]),basisn4m5)
      ↪ # The state in the Hilbert space basis we use
6   output_state=np.matmul(Ur,input_state.T)
7   print(leading_terms(output_state,0.99))
8   print(schmidt_rank_vector(output_state,basisn4m5,[1,1,1,1,1]))
```

The output subsystems are highly entangled and, in a typical run of this code, we obtain a Schmidt rank vector $[5, 5, 5, 5, 5]$. This is similar to what happens in the random evolutions used in boson sampling [9] and part of the reason why classical simulations struggle to simulate the output distribution.

However, the resulting states are a superposition of basically all the possible states in the basis (all the ways to put the $n$ photons in the $m$ modes). From the 70 states in the basis of the chosen example, depending on the concrete random matrix, there are around 40 terms which cumulatively have a probability of 0.9 of appearing in a measurement and usually more than 50 states are needed to explain the measurements in 99% of the cases. This kind of output superposition is not very practical for further use.

In our automated search for entanglement generating linear optical systems we would like to produce more compact output states. We give an example based on the heuristic that transformations producing cyclic rotations of the basis states are usually a prelude for entanglement generation [21].

Instead of searching for perfect cyclic rotations in a subset of all the possible inputs, we suppose that an approximation to a perfect rotation of all the basis states using `Toponogov` will produce an output which is, at the same time, compact and entangled. This has usually been the case in our experiments.

Notably, when approximating simple entangling unitary evolutions, we have found that, for simple inputs, the best approximation produces states with a small number of relevant terms, but the second best approximation or approximations with a larger matrix distance tend to take the input into a superposition of a larger number of states.

The presented example produces an state with two photons in three modes

$$|M\rangle = \frac{|011\rangle + |101\rangle + |110\rangle}{\sqrt{3}} \quad (31)$$

in a superposition where the empty mode is distributed in all the three possible positions. We have chosen the name $M$ state because the expression reminds of an inverted version of the $W$ states for three systems in a uniform superposition of only one system being excited.

The chosen configuration puts $n = 4$ photons into $m = 5$ modes. First, we approximate a rotation matrix in the $70 \times 70$ Hilbert space of the photons. Experimentally, we usually prefer states with a single photon in each mode, which are easier to produce and measure without photon number resolving detectors. We put those states close in the generated basis so that the rotation matrix we approximate would ideally move from one to another, at least for the first ones.

```
1  state_basis=subspace_basis(5,[1,0,1,1,1],[[0,1,1,1,1],
     [1,0,1,1,1],[1,1,0,1,1],[1,1,1,0,1],[1,1,1,1,0]])
2  Toponogov(file_input=False,U_input=RotMat(70,1),file_output=True,
     filename="Rotated",tries=10,m=5,n=4,acc_d=3,txt=False,
     acc_t=3,vec_base=state_basis)
```

In order to make the search efficient in time, we have reduced the convergence criterion of the approximation to $10^{-3}$. We tried 10 different initial random matrices, which took around 1.5 hours of computation in a 2.8 GHz core. We then took the best approximation, with a trace distance of 9.80 to the original matrix, and checked the output for the five possible input states where no mode holds more than one photon. We show the output for the most compact result.

```
3  U=read_matrix_from_txt('Rotated_toponogov_1') # Closest matrix
4  input_state=state_in_basis([[1,1,1,1,0]],np.array([1]),state_basis)
     # The state in the Hilbert space basis we use
5  output_state=np.matmul(U,input_state.T)
6  fidelity=0.99
7  shortstate,
     weights=state_leading_fidelity(output_state,state_basis,fidelity)
8  short=state_in_basis(shortstate,weights,state_basis)
```

For an input $|1,1,1,1,0\rangle$, the output $|\psi_{\text{out}}\rangle$ can be approximated by the state

$$|\psi_M\rangle = \alpha_1 |11011\rangle + \alpha_2 |11101\rangle + \alpha_3 |11110\rangle, \quad (32)$$

with $\alpha_1 = -0.11335 + 0.11385i$, $\alpha_2 = 0.35403 - 0.52643i$ and $\alpha_3 = 0.61096 - 0.43433i$ and $|\langle \psi_M | \psi_{\text{out}} \rangle|^2 > 0.99$.

The first two modes always carry a photon. They can be used as ancillary modes and measure them as a check and then work with the remaining three modes. We call the resulting state a partial $M$ state.

The output is a partial version with a somewhat large imbalance between terms, which appear with a probability $|\alpha_1|^2 = 0.0258$, $|\alpha_2|^2 = 0.4025$ and $|\alpha_3|^2 = 0.5619$. If we introduce two attenuators, one in the first mode (with a total transmission $T_1 = 0.046$) and one in the second mode (with a total transmission $T_2 = 0.716$), we obtain a balanced $M$ state with a 3.28% probability. This state can be used in experiments with postselection where we know that, if two photons are measured, we had the $M$ state at the beginning. The relative phases between the terms, if needed, can be corrected by choosing the correct combination of phase shifters for each of the three modes.

The unbalanced $M$ state in Equation (32) can be created in a setup with 10 beamsplitters and 15 phase shifters. The values of the required elements can be found using the functions `SfromU` and `Selements`. The elements for the matrix we have found are included with the code in the Examples folder.

*6.4. Decomposition of quasiunitary scattering matrices*

Since systems with loss are of great interest for further experiments of quantum computing with linear optics devices, we show two simple examples of the function `QuasiU`. They give the classical description of linear systems which can include losses and amplification.

**The first one** is the lossy beam splitter given by the $T$ transformation:

$$T = \frac{1}{2}\begin{pmatrix} 1 & -1 \\ -1 & 1 \end{pmatrix}, \tag{33}$$

which has already been analyzed in [39].

The only command required is

```
1    # Obtaining "T_dim2x2.txt"'s S-matrix.
2    QuasiU(file_input=True,filename="T_dim2x2",
         newfile=False,file_output=True)
```

`QuasiU` gives a lot of information about the experimental implementation of the transformation in Eq. (33). The algorithm decomposes any arbitrary complex matrix $M$ by using the Schur decomposition: generally, $M = UDW$, with $U$ and $W$ being unitary and $D$ a diagonal matrix.

Combining the latter result with the linear optic devices decomposition for unitary matrices such as $U$ and $W$ (see the use of `Selements` in the previous example), `QuasiU` returns the decomposition of $T$, which turns out to be a product of

$$U = U_D \cdot U_{T_{1,2}}(\theta = 0.785, \phi = 0.000) = \begin{pmatrix} -1.000 & 0.000 \\ 0.000 & 1.000 \end{pmatrix}\begin{pmatrix} 0.707 & -0.707 \\ 0.707 & 0.707 \end{pmatrix}, \tag{34}$$

$$D = \begin{pmatrix} 1.000 & 0.000 \\ 0.000 & 0.000 \end{pmatrix} \quad \text{and} \tag{35}$$

$$W = W_D \cdot W_{T_{1,2}}(\theta = 0.785, \phi = 0.000) = \begin{pmatrix} -1.000 & 0.000 \\ 0.000 & 1.000 \end{pmatrix}\begin{pmatrix} 0.707 & -0.707 \\ 0.707 & 0.707 \end{pmatrix}. \tag{36}$$

Since $D$ contains a value $d_{22} = 0 < 1$, there is loss in the second port. In order to give a complete description, we need a third ancilla mode (and the three matrices $U$, $D$, $W$ are padded with one extra dimension as a result).

This results in a straightforward quasiunitary system with loss in one mode via a virtual beam splitter incorporating the ancilla mode. The resulting scattering matrix for $T$ is:

$$S = \begin{pmatrix} 0.500 & -0.500 & 0.707 & 0.000 & 0.000 & 0.000 \\ -0.500 & 0.500 & 0.707 & 0.000 & 0.000 & 0.000 \\ -0.707 & -0.707 & 0.000 & 0.000 & 0.000 & 0.000 \\ 0.000 & 0.000 & 0.000 & 0.500 & -0.500 & 0.707 \\ 0.000 & 0.000 & 0.000 & -0.500 & 0.500 & 0.707 \\ 0.000 & 0.000 & 0.000 & -0.707 & -0.707 & 0.000 \end{pmatrix}. \tag{37}$$

This reconstructed matrix is stored in an output file when running `QuasiU`.

Non-diagonal matrix blocks are zero, which means no cross-interaction between the modes in each block, which are unitary. For such cases where there is no squeezing, the user can take out the first diagonal block and go back to a unitary representation of the device where we have needed to add ancillary modes to represent losses. The resulting scattering matrix is

$$S = \begin{pmatrix} 0.500 & -0.500 & 0.707 \\ -0.500 & 0.500 & 0.707 \\ -0.707 & -0.707 & 0.000 \end{pmatrix}. \tag{38}$$

As (38) is unitary, it is compatible with the other functions of `QOptCraft` and we can also compute its quantum evolution with `StoU` or find different experimental realizations of that lossy beam splitter with `Selements`.

In the **second example** we show the general method for arbitrary input complex matrices $M$. We can create new complex matrices either from the already known function `QuasiU` or the generator `RandM`.

In this example, we use the random matrices produced by `RandM`.

```
1    # We first generate the random matrix:
2    RandM(filename="M_dim2x3",N1=2,N2=3)
```

`RandM` generates a random $N_1 \times N_2$ complex matrix with elements drawn from a normal distribution in their real and imaginary parts. For our experiment, we picked a non-square 2x3 random complex matrix

$$M = \begin{pmatrix} 0.77 - 0.04i & -0.07 - 0.57i & 0.21 - 0.71i \\ 0.53 - 0.34i & 1.08 + 0.16i & -0.24 - 0.05i \end{pmatrix}. \tag{39}$$

The next command will be `QuasiU`'s execution:

```
3    # Obtains "M_dim2x3.txt"'s quasiunitary representation S.
4    QuasiU(file_input=True,filename="M_dim2x3",
5    newfile=False,file_output=True)
```

Looking at the output file of `QuasiU` with the decomposition, we see that, this time, the three matrices $U$, $D$, $W$ given by the Schur decomposition of (39) will be dimensionally different. All of them need to be expanded to quadratic $N$ x $N$ matrices, with $N$ being the highest of both $N_1, N_2$ (in this case, $N = 3$).

A quick gaze to the diagonal matrix $D$ allows us to detect gain on the first and second ports.

$$\begin{aligned} U &= U_D \cdot U_{T_{1,2}}(\theta = 0.93, \phi = 0.00) \\ &= \begin{pmatrix} -1.00 & 0.00 & 0.00 \\ 0.00 & -0.79 - 0.62i & 0.00 \\ 0.00 & 0.00 & 1.00 \end{pmatrix} \begin{pmatrix} 0.60 & -0.80 & 0.00 \\ 0.80 & 0.60 & 0.00 \\ 0.00 & 0.00 & 1.00 \end{pmatrix}, \end{aligned} \tag{40}$$

$$D = \begin{pmatrix} 1.37 & 0 & 0 \\ 0 & 1.12 & 0 \\ 0 & 0 & 1 \end{pmatrix} \quad \text{and} \tag{41}$$

$$\begin{aligned} W &= \left(W_{T_{1,2}^{even}}(\theta = 2.76, \phi = 0.58)\right)^{-1} \cdot \left(W_{T_{2,3}^{even}}(\theta = 2.41, \phi = 1.07)\right)^{-1} \\ &\quad \cdot W_D \cdot W_{T_{1,2}^{odd}}(\theta = 1.99, \phi = 6.03) \\ &= \begin{pmatrix} -0.78 - 0.51i & -0.37 & 0.00 \\ 0.31 + 0.20i & -0.93 & 0.00 \\ 0.00 & 0.00 & 1.00 \end{pmatrix}^{-1} \begin{pmatrix} 1.00 & 0.00 & 0.00 \\ 0.00 & -0.35 - 0.65i & -0.67 \\ 0.00 & 0.32 + 0.59i & -0.74 \end{pmatrix}^{-1} \\ &\quad \begin{pmatrix} -0.99 + 0.14i & 0.00 & 0.00 \\ 0.00 & -0.14 + 0.99i & 0.00 \\ 0.00 & 0.00 & -1.00 + 0.07i \end{pmatrix} \begin{pmatrix} -0.40 + 0.10i & -0.91 & 0.00 \\ 0.88 - 0.23i & -0.41 & 0.00 \\ 0.00 & 0.00 & 1.00 \end{pmatrix}. \end{aligned} \tag{42}$$

Gain devices, like parametric amplifiers, imply cross-interactions which prevent a unitary description. The quantum mechanical description of the oscillating EM field now requires explicit mention to both the creation and annihilation operators in the corresponding modes. These systems are called active, whereas loss-only cases akin to the previous example are known as passive.

Active systems do present no-null non-diagonal blocks in the scattering representation $S$ of $M$ as a consequence of cross-interaction. `QuasiU` reconstructs a valid quasiunitary matrix in an output file.

For our example, the file returns a scattering matrix $S$ which can be expressed in terms of two blocks $A$ and $B$. The complete description of the active linear system is given by a 10-dimensional quasiunitary matrix.

$$S = \begin{pmatrix} A & B^* \\ B & A^* \end{pmatrix}, \tag{43}$$

$$A = \begin{pmatrix} 0.77 - 0.04i & -0.07 - 0.57i & 0.21 - 0.71i & 0.00 & 0.00 \\ 0.53 - 0.34i & 1.08 + 0.16i & -0.24 - 0.05i & 0.00 & 0.00 \\ -0.07 - 0.61i & -0.04 + 0.27i & 0.74 - 0.05i & 0.00 & 0.00 \\ 0.00 & 0.00 & 0.00 & 1.37 & 0.00 \\ 0.00 & 0.00 & 0.00 & 0.00 & 1.12 \end{pmatrix},$$

$$B = \begin{pmatrix} 0.00 & 0.00 & 0.00 & -0.56 & 0.40 \\ 0.00 & 0.00 & 0.00 & -0.60 + 0.47i & -0.23 + 0.18i \\ 0.00 & 0.00 & 0.00 & 0.00 & 0.00 \\ -0.43 + 0.34i & -0.50 + 0.53i & 0.03 + 0.23i & 0.00 & 0.00 \\ 0.22 + 0.15i & -0.28 - 0.06i & 0.13 - 0.28i & 0.00 & 0.00 \end{pmatrix}.$$

## 7. Summary

We have presented the main functions and the theory behind the Python package `QOptCraft` for the automated design and analysis of quantum linear optical systems. This paper serves as a complement to the guide included with the software [1], where the reader can find the complete description of the function parameters as well as additional use examples. Here, we have explained the relationship between the different procedures and have related all the concepts to the groups and algebras that appear in the description of linear optical systems in a quantum setting.

The package can be used as a black box for experiment design or as a basic library to build more complex programs dealing with optical interferometers.

**Declaration of competing interest**

The authors declare that they have no known competing financial interests or personal relationships that could have appeared to influence the work reported in this paper.

**Data availability**

The code is available at https://github.tel.uva.es/juagar/qoptcraft.

**Acknowledgements**

The authors thank Alejandro Escorihuela Tomás for serving as a beta tester of the package. D. Gómez Aguado has been supported by the Spanish Government (Ministerio de Educación y Formación Profesional, Beca de Colaboración en Departamentos Universitarios). V. Gimeno has been partially supported by the Research Program of the Universitat Jaume I–Project UJI-B2018-35, as well as by the Spanish Government and FEDER grants PID2020-115930GA-I00 (MICINN) and MTM2017-84851-C2-2 (MINECO). J.J. Moyano-Fernández was partially supported by MCIN/AEI/10.13039/501100011033 and by "ERDF A way of making Europe", grants PGC2018-096446-B-C22 and RED2018-102583-T, as well as by Universitat Jaume I, grant UJI-B2021-02. J.C. Garcia-Escartin has been funded by the Spanish Government and FEDER grant PID2020-119418GB-I00 (MICINN) and Junta de Castilla y León (project VA296P18).